\documentclass[apj]{emulateapj}
\usepackage{graphicx,txfonts,amssymb,amsfonts,natbib,times,hyperref}
\usepackage{url}

\begin{document}
\submitted{Submitted to The Astrophysical Journal Letters}

\newcommand{\boot}{Bo\"otes}
\newcommand{\kms}{~km~s$^{-1}$}
\newcommand{\logh}{$+5\log h_{70}$}
\newcommand{\ho}{~}
\newcommand{\hi}{$h^{-1}_{70}$~}
\newcommand{\msun}{M$_\odot$}
\newcommand{\degtwo}{deg$^2$}
\newcommand \sbu {mag arcsec$^{-2}$}
\newcommand \halpha{H$_\alpha$}
\newcommand \hbeta{H$_\beta$}
\newcommand \hdelta{H$_\delta$}
\newcommand \firstpaper {(Paper I)}
\newcommand{\spitzer}{\textit{Spitzer}}
\newcommand{\wise}{\textit{WISE}}
\newcommand{\allwise}{All\textit{WISE}}
\newcommand{\galex}{{\it GALEX}}
\newcommand{\herschel}{\textit{Herschel}}
\newcommand{\planck}{\textit{Planck}}
\newcommand{\hst}{\textit{HST}}
\newcommand{\chandra}{\textit{Chandra}}
\newcommand{\irac}{IRAC}
\newcommand{\irs}{IRS}
\newcommand{\mips}{MIPS}
\newcommand{\wfc}{WFC3}
\newcommand{\acs}{ACS}
\newcommand{\kcorrect}{\texttt{kcorrect}}
\newcommand{\galfit}{\texttt{GALFIT}}
\newcommand{\multidrizzle}{\texttt{MultiDrizzle}}
\newcommand{\hyperz}{\texttt{HyperZ}}
\newcommand{\ergflux}{~erg~s$^{-1}$~cm$^{-2}$}
\newcommand{\micr}{$\mu$m}
\newcommand{\mfive}{M$_{500}$}
\newcommand{\mfivec}{M$_{500c}$}
\newcommand{\mtwo}{M$_{200c}$}
\newcommand{\mtwom}{M$_{200m}$}
\newcommand{\mtwoc}{M$_{200c}$}
\newcommand{\rfive}{$r_{500}$}
\newcommand{\rfivec}{$r_{500c}$}
\newcommand{\rtwo}{$r_{200}$}
\newcommand{\chone}{[$3.6${}]}
\newcommand{\chtwo}{[$4.5${}]}
\newcommand{\wone}{W1{}}
\newcommand{\wtwo}{W2{}}
\newcommand{\madcows}{{MaDCoWS}}
\newcommand{\target}{{MOO J1142+1527}}
\newcommand{\idcscluster}{{IDCS J1426.5+3508}}
\newcommand{\xmmgroup}{{XMM-LSS J02182-05102}}
\newcommand{\xdcpcluster}{{XDCP J004.0-2033}}
\newcommand{\bootarea}{{8.82}}
\newcommand{\sz}{Sunyaev-Zel'dovich}
\newcommand{\lcdm}{$\Lambda$CDM}
\newcommand\mstar{$M^\ast$}
\newcommand\lstar{$L^\ast$}
\newcommand\targz{1.19}

\renewcommand{\d}{\mathrm{d}}

\title{The Massive and Distant Clusters of \wise\ Survey :\\ \target,
A $10^{15}$ \msun\ Galaxy Cluster at $z=1.19$}
\shorttitle{\target: A $10^{15}$ \msun\ Galaxy Cluster at $z=1.19$}

\author{ Anthony H. Gonzalez\altaffilmark{1},
  Bandon Decker\altaffilmark{2},
  Mark Brodwin\altaffilmark{2},
  Peter R. M. Eisenhardt\altaffilmark{3},
  Daniel P. Marrone\altaffilmark{4},
  S. A. Stanford\altaffilmark{5,6},  	
  Daniel Stern\altaffilmark{3}, 	
  Dominika Wylezalek\altaffilmark{7},	
  Greg Aldering\altaffilmark{8},	
  Zubair Abdulla\altaffilmark{9,10},	
  Kyle Boone\altaffilmark{8,11},	
  John Carlstrom\altaffilmark{9,10},	
  Parker Fagrelius\altaffilmark{8,11},	
  Daniel P. Gettings\altaffilmark{1},	
  Christopher H. Greer\altaffilmark{4},	
  Brian Hayden\altaffilmark{8,12},	
  Erik M. Leitch\altaffilmark{9,10},	
  Yen-Ting Lin\altaffilmark{13},	
  Adam B. Mantz\altaffilmark{14,15},	
  Stephen Muchovej\altaffilmark{16,17},	
  Saul Perlmutter\altaffilmark{8,11}, and
  Gregory R. Zeimann\altaffilmark{18}	
  }

\altaffiltext{1}{Department of Astronomy, University of Florida, Gainesville, FL 32611-2055}
\altaffiltext{2}{Department of Physics and Astronomy, University of Missouri, 5110 Rockhill Road, Kansas City, MO, 64110}
\altaffiltext{3}{Jet Propulsion Laboratory, California Institute of Technology, Pasadena, CA 91109}
\altaffiltext{4}{Steward Observatory, University of Arizona, Tucson, AZ 85121}
\altaffiltext{5}{Department of Physics, University of California, One Shields Avenue, Davis, CA 95616}
\altaffiltext{6}{Institute of Geophysics and Planetary Physics, Lawrence Livermore National Laboratory, Livermore, CA 94550}
\altaffiltext{7}{Department of Physics and Astronomy, Johns Hopkins University, Baltimore, MD 21218}
\altaffiltext{8}{Physics Division, Lawrence Berkeley National Laboratory, 1 Cyclotron Road, Berkeley, CA 94720}
\altaffiltext{9}{Department of Astronomy and Astrophysics, University of Chicago, Chicago, IL 60637}
\altaffiltext{10}{Kavli Institute for Cosmological Physics, University of Chicago, Chicago, IL 60637}
\altaffiltext{11}{Department of Physics, University of California Berkeley, Berkeley, CA 94720}
\altaffiltext{12}{Space Sciences Lab, University of California Berkeley, 7 Gauss Way, Berkeley, CA 94720}
\altaffiltext{13}{Institute of Astronomy and Astrophysics, Academica Sinica, Taipei, Taiwan}
\altaffiltext{14}{Kavli Institute for Particle Astrophysics and Cosmology, Stanford University, 452 Lomita Mall, Stanford, CA 94305}
\altaffiltext{15}{Department of Physics, Stanford University, 382 Via Pueblo Mall, Stanford, CA 94305}
\altaffiltext{16}{California Institute of Technology, Owens Valley Radio Observatory, Big Pine, CA 93513}
\altaffiltext{17}{California Institute of Technology, Department of Astronomy, Pasadena, CA 91125}
\altaffiltext{18}{Department of Astronomy and Astrophysics, Pennsylvania State University, University Park, Pennsylvania 16802}

\begin{abstract}
  We present confirmation of the cluster \target, a massive galaxy
  cluster discovered as part of the Massive and Distant Clusters of
  \wise\ Survey.  The cluster is confirmed to lie at $z=\targz$, and
  using the Combined Array for Research in Millimeter-wave Astronomy
  we robustly detect the \sz\ (SZ) decrement at 13.2$\sigma$.  The SZ data
  imply a mass of \mtwom$=(1.1\pm0.2)\times10^{15}$ \msun, making
  \target\ the most massive galaxy cluster known at $z>1.15$ and the
  second most massive cluster known at $z>1$. For a standard
  $\Lambda$CDM cosmology it is further expected to be one of the $\sim
  5$ most massive clusters expected to exist at $z\ge1.19$ over the
  entire sky. Our ongoing \spitzer\ program targeting $\sim1750$
  additional candidate clusters will identify 
  comparably rich galaxy clusters over the full extragalactic sky.
\end{abstract}


\section{Introduction}
In the past few years we have entered a new era of wide-area surveys
capable of detecting galaxy clusters at $z>1$. The previous generation
of high-redshift cluster searches was the first to yield large samples
of galaxy clusters at this epoch
\citep[e.g.][]{gladders2005,eisenhardt2008,muzzin2009,fassbender2011};
however, these programs typically probed less than 100
\degtwo. Consequently, while these surveys have been effective in
generating statistical samples of distant galaxy clusters, they have
lacked the comoving volume to discover significant numbers of massive
clusters (\mfivec$\ga 3\times10^{14}$ \msun).
The high mass tail of the galaxy cluster population is of significant
interest for both galaxy evolution and cosmology. One open question is
the extent to which the star formation, active galactic nuclei (AGN)
activity, and assembly histories of cluster galaxies depend upon the
mass of the cluster in which they reside
\cite[e.g.][]{brodwin2013,ehlert2015,ma2015}. For this science,
samples of high-mass clusters close to the epochs of assembly and star
formation, coupled with existing lower mass samples, provide the
necessary dynamical range to quantitatively address this question. For
cosmology, massive, high-redshift clusters remain competitive probes
of dark energy via a number of methods 
\citep[e.g.][]{allen2011}, including evolution in the cluster mass function
\citep{vikhlinin2009,bocquet2015}, the clustering of galaxy clusters
\citep[e.g.][]{sereno2015}, and through application of the $f_{gas}$
test \citep{mantz2014}. The high mass tail of the galaxy cluster mass
function is also a sensitive indicator of primordial non-Gaussianity
\citep{chen2010,williamson2011,shadera2013}.

\begin{figure*}
\epsscale{1.15}
\begin{center}
\includegraphics[width=8cm]{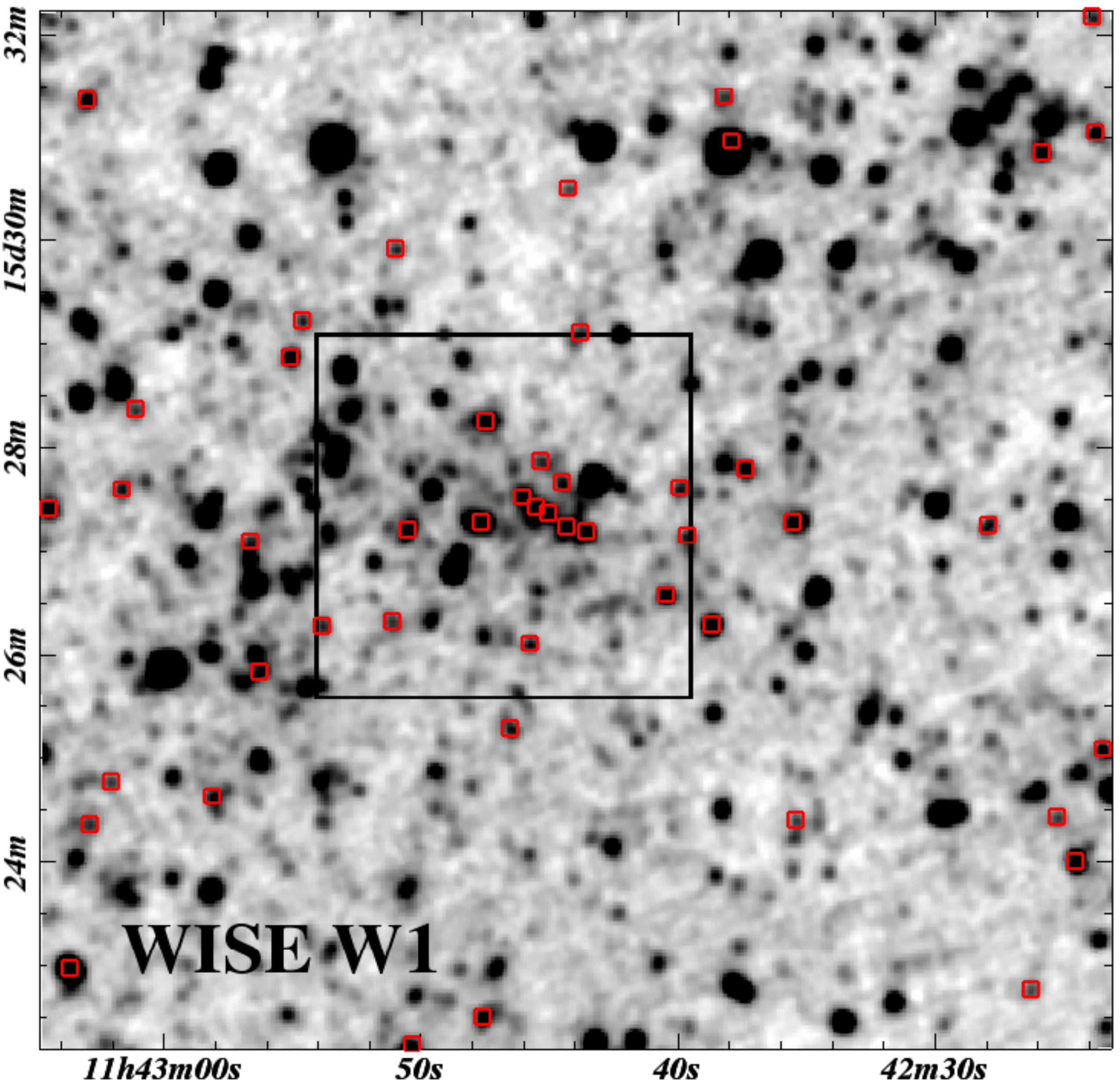}
\includegraphics[width=8cm]{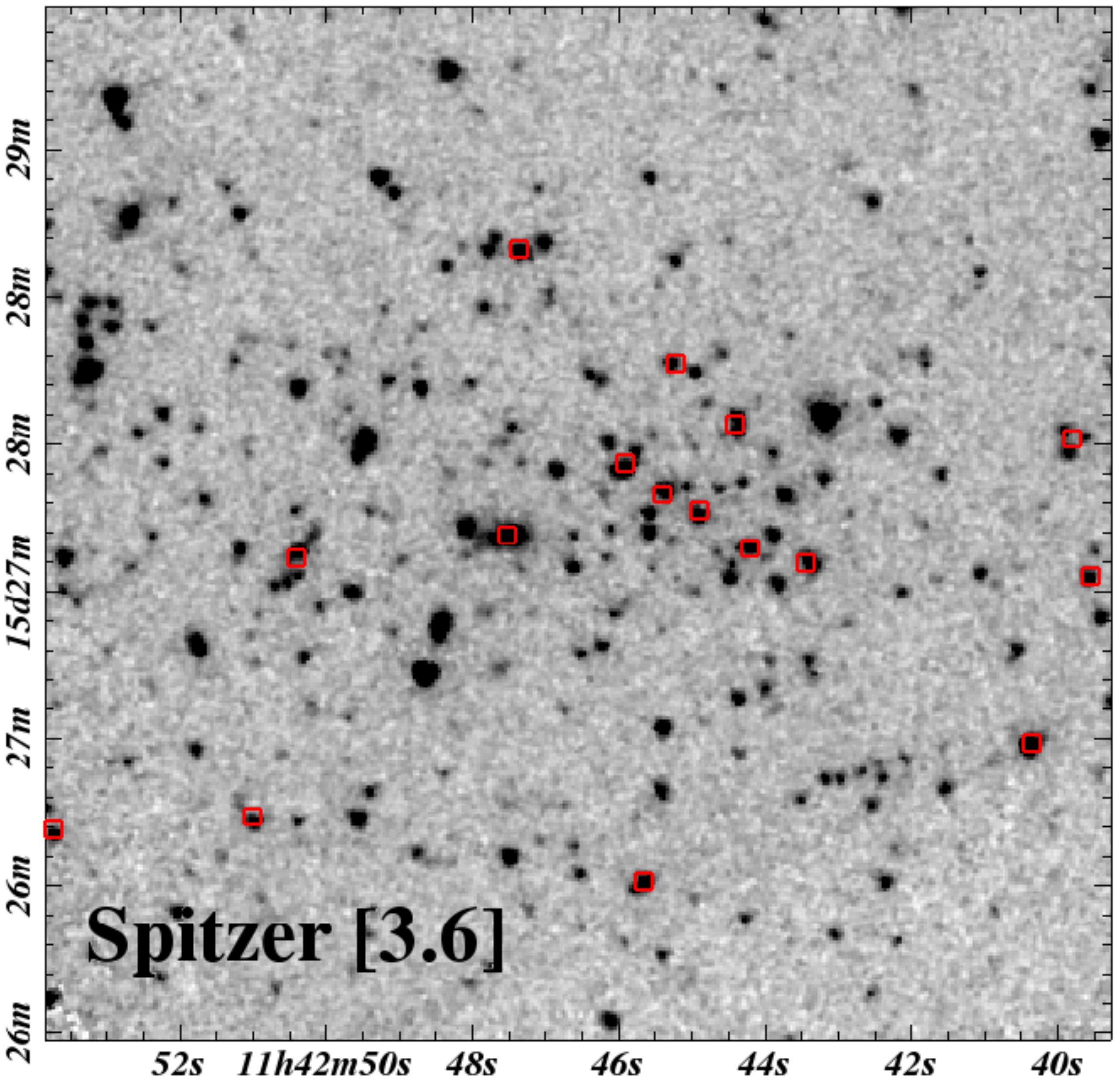}
\end{center}
\caption{ The left panel shows the $10^\prime\times10^\prime$ \wone\
  cutout of \target\ from the \allwise\ data release. The black box denotes the
  $3.5^\prime\times3.5^\prime$ region centered on the cluster for which
  we show the corresponding \spitzer\ \chone\ follow-up observation on
  the right. In both panels the red points denote the locations of
  individual \wise\ sources that pass the color, magnitude, and
  quality cuts as candidate $z\ga0.75$ galaxies in the \madcows\
  search. In several cases individual \wise\ sources are resolved into multiple
galaxies in the higher resolution \spitzer/\irac\ images.}
\label{fig:wise}
\end{figure*}

In the last several years, the South Pole Telescope (SPT) and Atacama
Cosmology Telescope (ACT) have each completed wide-area millimeter
surveys to identify galaxy clusters via the \sz\ effect, publishing
cluster catalogs drawn from 2500 \degtwo\ for the SPT survey
\citep{bleem2015} and 504 \degtwo\ for the ACT survey
\citep{hasselfield2013}. Together, these programs have published
nearly 50 massive clusters at $z>1$. The upcoming generation of
optical, galaxy-based cluster searches will also extend into the
wide-area, high-redshift region of parameter space, complementing
these millimeter surveys.  When complete, the Dark Energy Survey
\citep{flaugher2005,sanchez2010} is expected to result in a
cluster catalog extending to $z\sim1$, covering a $\sim5000$ \degtwo\
footprint that includes much of the SPT and ACT survey areas.

The Massive and Distant Clusters of \wise\ Survey (\madcows), which is
designed to detect the most massive galaxy clusters at $z\approx1$,
offers the largest survey area among current high-redshift cluster
searches.  The first phase of \madcows\ covered $\sim10,000$ \degtwo\
within the SDSS footprint; subsequent phases of the program are now
extending the search over the full extragalactic sky. In previous
papers we presented the first cluster discovered in this survey
\citep{gettings2012}, the redshift distribution of the first 20
clusters \citep[$0.75<z<1.3$,][]{stanford2014}, and \sz\ masses for
five clusters \citep{brodwin2015}.  In this paper we present the
discovery and confirmation of the most massive cluster yet identified
within the \madcows\ catalog, which is among the $\sim$5 most massive
clusters expected to exist over the entire sky at $z\ga1.19$.
Throughout the paper we use Vega-based magnitudes and assume a WMAP9
cosmology \citep[$H_0=69.7$\kms, $\Omega_m=0.2821$,
$\Omega_\Lambda=0.7181$, $\sigma_8=0.817$,
$n_s=0.9646$;][]{hinshaw2013} unless otherwise specified. 
For cluster masses and radii we include a {\it c} or {\it m} subscript
to denote whether the values are relative to the critical or mean
density.

\section{Discovery of \target}
\madcows\ is a \wise-based \citep{wright2010} search for galaxy clusters at $z\simeq1$
that employs color and magnitude selection to identify massive
galaxies at $z\ga0.75$, and then uses a wavelet technique to detect
galaxy overdensities. A key element of this search approach is the
combination of the \wise\ data with uniform optical photometry.  The
initial detection of Massive Overdense Object (MOO)
J1142+1527 
used the \wise\ All-Sky Data Release
\citep{cutri2012}
and SDSS DR8 \citep{dr8_paper} to identify candidates within the
footprint of the SDSS.  In this \wise+SDSS \madcows\ search, \target\
was identified as one of the 200 highest significance cluster
candidates.

We have subsequently refined the search algorithm and transitioned to
use of the \allwise\ Data Release \citep{cutri2013}.
A detailed description of the \madcows\ survey and detection algorithm
will be provided in a forthcoming paper.  Briefly, cluster candidates
in the current \allwise+SDSS search were detected as overdensities of
sources with \wone$< 16.9$, \wone--\wtwo$>0.2$, and $i_{AB}>21.3$.
\target\ remains the twenty-seventh highest significance candidate in
this more recent, refined version of the catalog,
with a position
$(\alpha,\delta)=$(11:42:43.9,15:27:07).  In the left panel of Figure
\ref{fig:wise} we show a \wise\ \chone\ cutout of the cluster
field. The red squares in this panel denote the galaxies that passed
the color, magnitude, and quality cuts in this search, highlighting
the detected overdensity.  Because the \wise\ magnitude limit and
blending of sources in the \wise\ data result in detection
significance being a high scatter richness measure, we have obtained
\spitzer/IRAC observations to determine more robust richness
estimates.

\section{{\it Spitzer} Richness and Color-Magnitude Diagram}
In \spitzer\ Cycle 9 we were awarded 37.9 hours to obtain IRAC 3.6\micr\ and
4.5\micr\ imaging of the 200 highest significance overdensities from
our All-Sky search (Program ID 90177; PI Gonzalez).  For each cluster
the total exposure times were 180~s in each band, obtained using 30~s
frame times and 6 positions in a medium scale cycling dither
pattern. This exposure time was designed to reach a nominal 5$\sigma$
depth of 6 $\mu$Jy (18.7 mag) at \chtwo, which is sufficient to
identify galaxies more than one magnitude below \lstar\ up to
$z\simeq1.5$.

We reduced and mosaicked the basic calibrated data using the MOPEX
package \citep{makovoz2005} and resampled to a pixel scale of
$0\farcs6$. The MOPEX outlier (e.g., cosmic ray, bad pixel) rejection
was optimized for the regions of deepest coverage in the center of the
maps corresponding to the position of the \madcows\ detection.

We ran SExtractor \citep{bertin1996} in dual image mode for source
detection and photometry, using the \chtwo\ frame as the detection
image and adopting IRAC-optimized SExtractor parameters from
\citet{lacy2005}.  Flux densities were measured in 4$^{\prime\prime}$
diameter apertures.  Following \citet{wylezalek2013}, we then applied
aperture corrections to the 
\chone\ and \chtwo\ flux densities (factors of 1.42 and 1.45, respectively). 
We determined a
95\% completeness limit of 10 $\mu$Jy, corresponding to limiting
magnitude of \chtwo\ $= 18.2$.  This completeness limit is adopted as the flux
density cut in all subsequent analysis.  We show the central
$3.5\arcmin\times3.5\arcmin$ of the \irac\ \chone\ image in the right
panel of Figure \ref{fig:wise}. As in the left panel, the red squares
denote the positions of \wise\ sources that contributed to detection
of the cluster. In some cases, the initial \wise\ source resolves into
multiple galaxies with \irac.

To prioritize follow-up of our Cycle 9 IRAC targets, we defined a
simple richness estimator based upon the overdensity of galaxies with
red \chone\ $-$ \chtwo\ color within a fixed angular radius.
Specifically, we defined the richness as the number of galaxies with
\chone\ $-$ \chtwo\ $>0.1$ within 1$^\prime$ of the cluster position
measured in the \madcows\ search. We note that this \irac\ color is
relatively insensitive to current star formation, selecting both
passive and star-forming galaxies in distant clusters.  By this
measure, \target\ has a richness of 64, which is the ninth highest
among the 200 Cycle 9 targets.

\begin{figure}
\epsscale{1.15}
\plotone{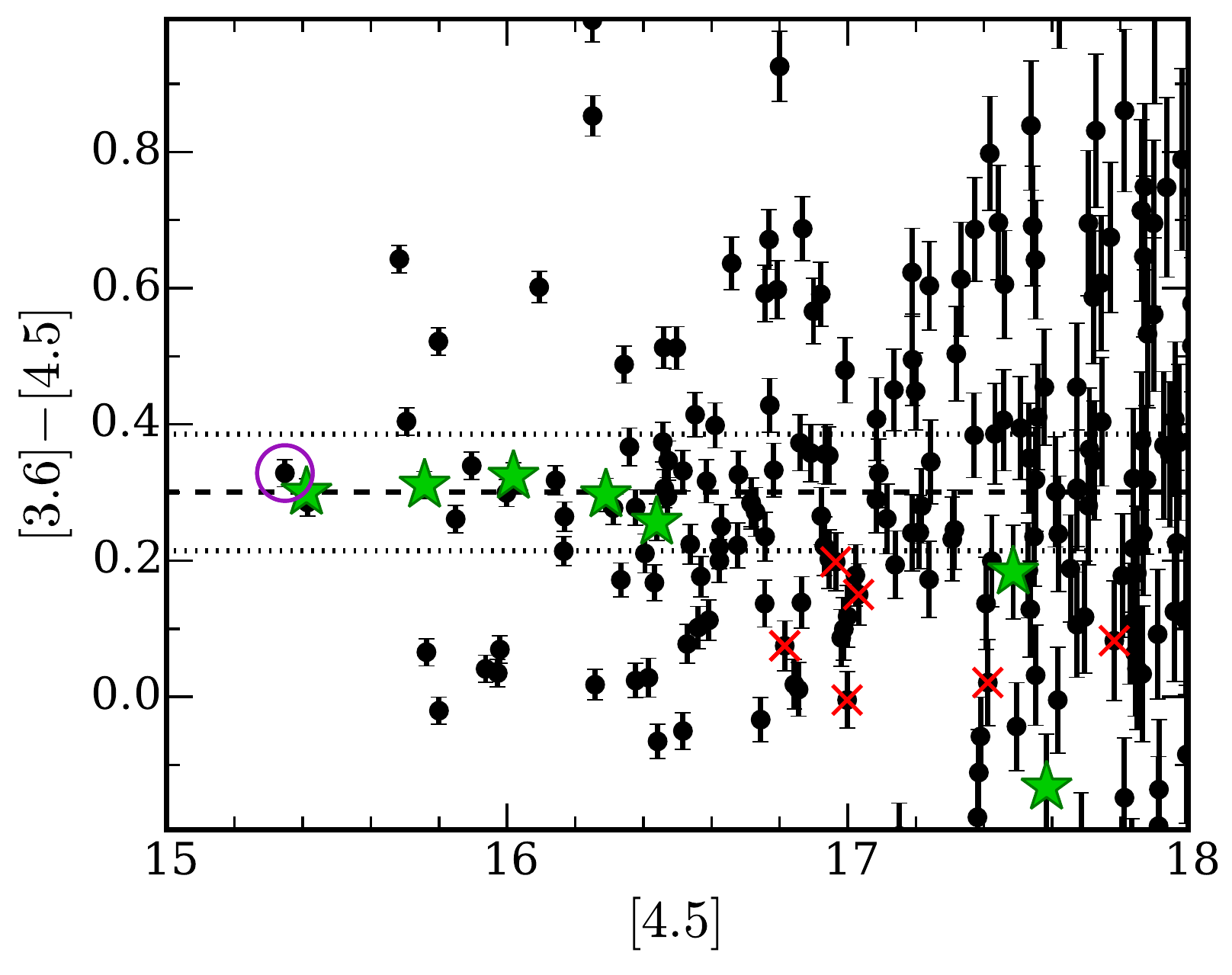}
\caption{ The 
 \spitzer\ \chone\ $-$ \chtwo\ color-magnitude diagram for \target. The black filled circles 
represent galaxies that lie within 1$^{\prime}$ of the SZ centroid. 
Solid green stars denote quality A and B spectroscopic members, while red crosses indicate foreground and background objects listed in Table \ref{tab:redshifts}. The open purple circle denotes the galaxy corresponding to the NVSS radio point source (\textsection{\ref{sec:sz}}). The dashed black line is the expected color from a \citet{bc03} model of a passively evolving, solar metallicity \lstar\ galaxy with a formation redshift $z_f=3$ at $z=1.19$; dotted lines indicate the equivalent expected colors for $z=1.09$ (lower) and $z=1.29$ (upper).} 
\label{fig:cmd}
\end{figure}

Figure \ref{fig:cmd} shows the \spitzer\ \chone\ $-$ \chtwo\
color-magnitude diagram for 
galaxies that lie within 1\arcmin\ of the SZ centroid (see \textsection{\ref{sec:sz}}).  These galaxies correspond to the central overdensity of red
sources shown in Figure \ref{fig:wise}.  The median color of these
galaxies is \chone\ $-$ \chtwo\ $=0.3$, which for a \citet{bc03}
passively evolving stellar population corresponds to a galaxy at
$z\simeq1.2$. We also highlight the spectroscopically confirmed
members (green stars), four of which lie within 1\arcmin of the SZ
centroid, and the non-members (red crosses), which are described in
greater detail in the next section.

\subsection{Redshift Determination}

We used Gemini-North and the W. M. Keck Observatories to obtain spectroscopic
confirmation of \target.  Optical pre-imaging for \target\ was
obtained with the Gemini Multi-Object Spectrograph (GMOS) on
Gemini-North as part of progam GN-2013A-Q-44 (PI Brodwin).  We
acquired 900~s exposures in the $r$- and $z$-bands, sufficient to
detect cluster galaxies below \lstar\ at the cluster redshift. Image
quality was $0\farcs68$ for $r$ and $0\farcs76$ for $z$.  For all
spectroscopic programs we designed slit masks using the Gemini
$rz$-band catalogs to identify potential cluster members.  We used the
red sequence to select the primary targets, weighting by
cluster-centric radius, and then filling in the masks with other
galaxies at larger radii.

\begin{deluxetable}{lllcc}
\tabletypesize{\scriptsize}
\tablecaption{Spectroscopic Redshifts}
\tablehead{
\colhead{$\alpha$ } & \colhead{$\delta$} &  \colhead{$z$} & \colhead{Quality} & \colhead{Features} 
 }
\startdata
\multicolumn{5}{c}{Spectroscopic Members}\\
\hline
11:42:40.04 & +15:26:28.1 & 1.2007 & A & \hbeta,[OIII]$\lambda$4959,5007  \\
11:42:40.31 & +15:26:28.4 & 1.20 & B & D4000 \\ 
11:42:42.14 & +15:26:59.9 & 1.19   & B & D4000 \\ 
11:42:43.36 & +15:27:05.2 & 1.19   & B & D4000 \\ 
11:42:43.83 & +15:27:01.6 & 1.179 & A & Ca HK,\hdelta  \\
11:42:45.82 & +15:27:25.0 & 1.19 & B & D4000 \\	
11:42:49.62 & +15:26:59.8 & 1.1715 & B & \hbeta,[OIII]$\lambda$5007  \\ 
11:42:54.09 & +15:26:54.3 & 1.183 & B & \hbeta,[OIII]$\lambda$5007 \\  
\hline\hline
\multicolumn{5}{c}{Foreground/Background Objects}\\
\hline
11:42:41.29 & +15:27:59.4 & 0.7221& A & \hbeta,[OIII]$\lambda$5007 \\ 
11:42:42.30 & +15:26:00.6 & 0.92 & B & D4000   \\ 
11:42:42.40 & +15:26:22.1 & 1.054 & B & Ca HK \\ 
11:42:42.69 & +15:26:23.5 & 1.2342 & B & [OII]$\lambda$3727 \\ 
11:42:44.07 & +15:27:02.4 & 1.2401 & A & [OIII]$\lambda$4959,5007 \\  
11:42:44.94 & +15:27:44.7 & 0.93  & B & Ca HK \\
11:42:45.22 & +15:28:07.5 & 0.93  & B & Ca HK  
\enddata
\label{tab:redshifts}
\tablecomments{This Table includes all spectroscopic redshifts for objects within 2\arcmin\ of the SZ centroid for the cluster. 
}
\end{deluxetable}

We obtained Gemini GMOS spectroscopy in queue mode on UT 2013 July 02
and UT 2014 March 07, using $1\farcs0 $ slit widths, the R400 grating,
and the RG610 filter.  Three sets of nod and shuffle sequences were
completed at each of two central wavelength settings (8100 and 8200
$\AA$).  For each nod and shuffle sequence we used $\pm 0\farcs75$
nods, with 9 cycles of 60~s exposures, yielding a total on-source
exposure time of 6480~s.  The seeing ranged between $0\farcs6$ and
$0\farcs9$ range.  We reduced the spectra using standard routines in
the Gemini IRAF package.

\begin{figure*}
\begin{center}
\epsscale{1.15}
\plotone{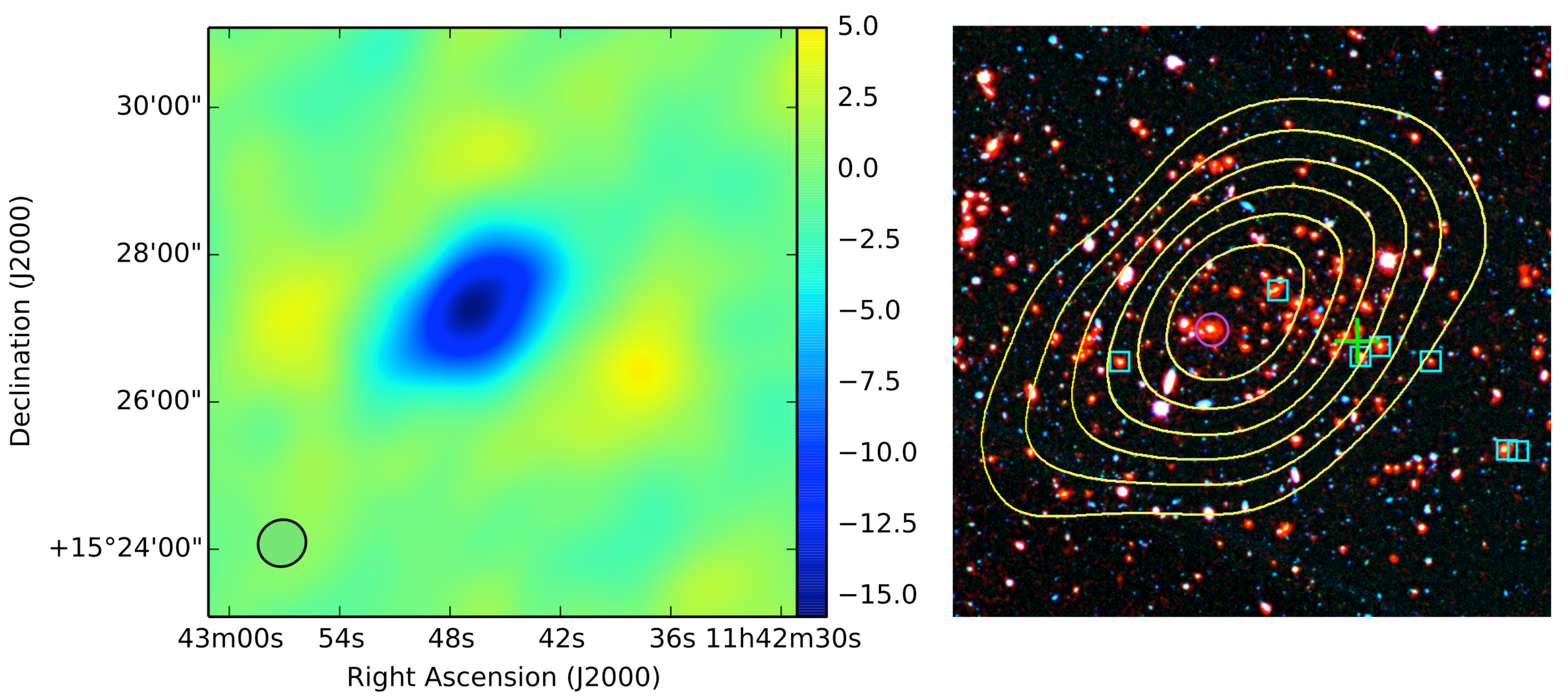}
\end{center}

\caption{Left: CARMA SZ map of the cluster covering an
  $8^\prime\times8^\prime$ field, where the scale denotes the significance
  of the decrement. The black ellipsoid shows the size of the synthesized beam,  which has major and minor axes of 39.7$\arcsec$ and 37.6$\arcsec$. Right: Composite $rz[3.6]$ image of \target, covering
  the central $3.5^\prime\times3.5^\prime$, with the SZ contours
  overlaid. The outer contour corresponds to 3$\sigma$, with
  subsequent contours incrementing by 2$\sigma$. The plus sign
  indicates the original \madcows\ position, while small squares
  denote spectroscopic cluster members. The purple circle marks NVSS J114247+152711.}
\label{fig:sz}
\end{figure*}

We subsequently obtained DEIMOS and MOSFIRE spectroscopy at the Keck
Observatory on UT 2015 May 12 and UT 2015 June 22, respectively.  For
DEIMOS, the masks were designed with $1\farcs1$ width slitlets having
a minimum length of 5\arcsec. In addition to the standard target
selection criteria, for these masks the \wise\ \wone$-$\wtwo\ color
was used to prioritize targets at large radii. Observations for two
masks were obtained under cloudy conditions with typical seeing of
$0\farcs8$. Four exposures of 1800~s each were obtained on the first
mask, and three exposures of 1500~s on the second mask. Both masks
used the 600ZD grating with the GG495 filter. We reduced these DEIMOS
spectra using the DEEP2 pipeline \citep{cooper2012,newman2013}.

For MOSFIRE, the configurable slit unit was configured for 32 objects,
along with five alignment stars.  We chose to use the $Y$ bandpass
because it covers a spectral range of $\sim$9900 to $\sim$11200 $\AA$,
which encompasses strong rest frame optical emission lines such as [O
III]$\lambda 4959, 5007$ at the probable cluster redshift.  The
MOSFIRE spectra were obtained using an ABA$^\prime$B$^\prime$ dither
pattern with 120~s exposures and multiple correlated double sampling
(MCDS), in the MCDS 16 readout mode.  The total integration time was
5760~s.  Conditions during the observations were excellent, with
seeing measured at $\sim$0\farcs5.  
MOSFIRE spectra were reduced using the standard MOSFIRE data reduction
pipeline.\footnote{\url{https://keck-datareductionpipelines.github.io/MosfireDRP/}}

The redshift determinations from the combination of Gemini/GMOS,
Keck/DEIMOS, and Keck/MOSFIRE spectroscopy are shown in Table
\ref{tab:redshifts} for all galaxies that lie within 2\arcmin\ ($\sim
1$ Mpc) of the cluster center.  We assigned redshifts a quality of A
if there are multiple obvious features associated with the same rest
frame redshift.  Quality B was assigned to redshifts that satisfy one
of the following: one and only one emission line is present and is
highly likely to be [O II]$\lambda$3727 given the observed wavelength
range of the spectra, an obvious 4000 $\AA$ feature is seen but no
other features, or Ca H$+$K absorption lines are clearly identified.
We determined the mean redshift using the \citet{ruel2014} python
implementation of the \citet{beers1990} biweight estimator.  The
resulting cluster redshift estimate is $z=1.188^{+0.002}_{-0.005}$,
with the uncertainty derived via bootstrap resampling.  The eight
galaxies listed as spectroscopic members in Table \ref{tab:redshifts}
were those that are retained as members by the redshift estimation
code after sigma-clipping.

\section{The Sunyaev-Zel'dovich Decrement and Derived Mass}
\label{sec:sz}
\target\ was observed with the Combined Array for Research in
Millimeter-wave Astronomy (CARMA)\footnote{\url{http://www.mmarray.org}}
for approximately 5 hours on-source beginning on UT 2014 July 03.
The data are centered around a frequency of 31~GHz. For these observations the array was in its most compact ``E+SH'' configuration. All 23 antennas were correlated across 2~GHz of bandwidth using the CARMA ``spectral line'' correlator. To maximize sensitivity to the SZ signal, the ``wideband'' correlator processed 7.5~GHz of bandwidth for the innermost eight 6.1-meter antennas. CARMA is optimized for the detection of distant clusters via their SZ signatures in this array and correlator configuration. The data from these 
baselines achieve a sensitivity of 1.2~mJy per $\sim50\arcsec\times90\arcsec$ beam.
The gain calibrator J1224+213 was observed for 3 minutes between 15-minute target observations, and the absolute calibration is derived from Mars via the model of \citet{rudy1987}. Figure \ref{fig:sz} (left) shows a CLEAN-deconvolved \citep{hogbom1974} image of the cluster using all baselines with a Gaussian taper to 10\% at 4~k$\lambda$, after removal of a point source (see below).

Cluster properties were determined using a Markov Chain Monte
Carlo method to simultaneously fit an \citet{arnaud2010} pressure
profile model and point source models to the unflagged data. The
single point source in the field, NVSS J114247+152711
\citep{condon1998}, was located using the higher-resolution, long
baseline data from the spectral line correlator. This point source was
found to have a flux density of 3.4 mJy, and is coincident with the
brightest candidate cluster member (purple circle in
Figures \ref{fig:cmd} \& \ref{fig:sz}), but was not targeted in our spectroscopic program.  A second model, consisting of only the point
source but no cluster, was also fit to the data.  From comparing the
goodness of the two fits, the cluster detection significance is 13.2
$\sigma$.  
The centroid of the SZ decrement is located at
$(\alpha,\delta)=$(11:42:46.6, +15:27:15), with uncertainties
$(\sigma_\alpha,\sigma_\delta)=(4\farcs4,3\farcs0)$.  The SZ centroid
and \wise\ position, which are separated by 41\arcsec , bracket the
peak of the galaxy distribution.

The combined fit of cluster and point source models gives the
spherically integrated Comptonization parameter, $Y_{500c} = (9.7 \pm
1.3) \times 10^{-5}$ Mpc$^2$.  We used the \citet{andersson2011}
\mfivec --$Y_{500}$ scaling relation to determine mutually consistent
values of \mfivec\ and \rfivec\ and the associated uncertainties, where \mfivec\ $ = (4\pi r_{500}^3/3)
\times (500 \rho_c)$.   This procedure results in a cluster mass and
radius of \mfivec\ $= (6.0 \pm 0.9) \times 10^{14}$ \msun\ and
\rfivec\ $= 0.83 \pm 0.04$ Mpc, respectively. 
The quoted uncertainties are derived by combining in quadrature the propagated uncertainty and a 12\% intrinsic scatter in \mfivec\ at fixed $Y_{500c}$ from
\citet{andersson2011}.
For the \citet{duffy2008} mass -- concentration relation, the derived mass corresponds to
\mtwoc$=(9.9\pm1.5) \times 10^{14}$ \msun, or
\mtwom$=(1.1\pm0.2)\times 10^{15}$ \msun.

\section{Discussion and Summary}
\label{sec:summary}

\begin{figure}
\epsscale{1.15}
\plotone{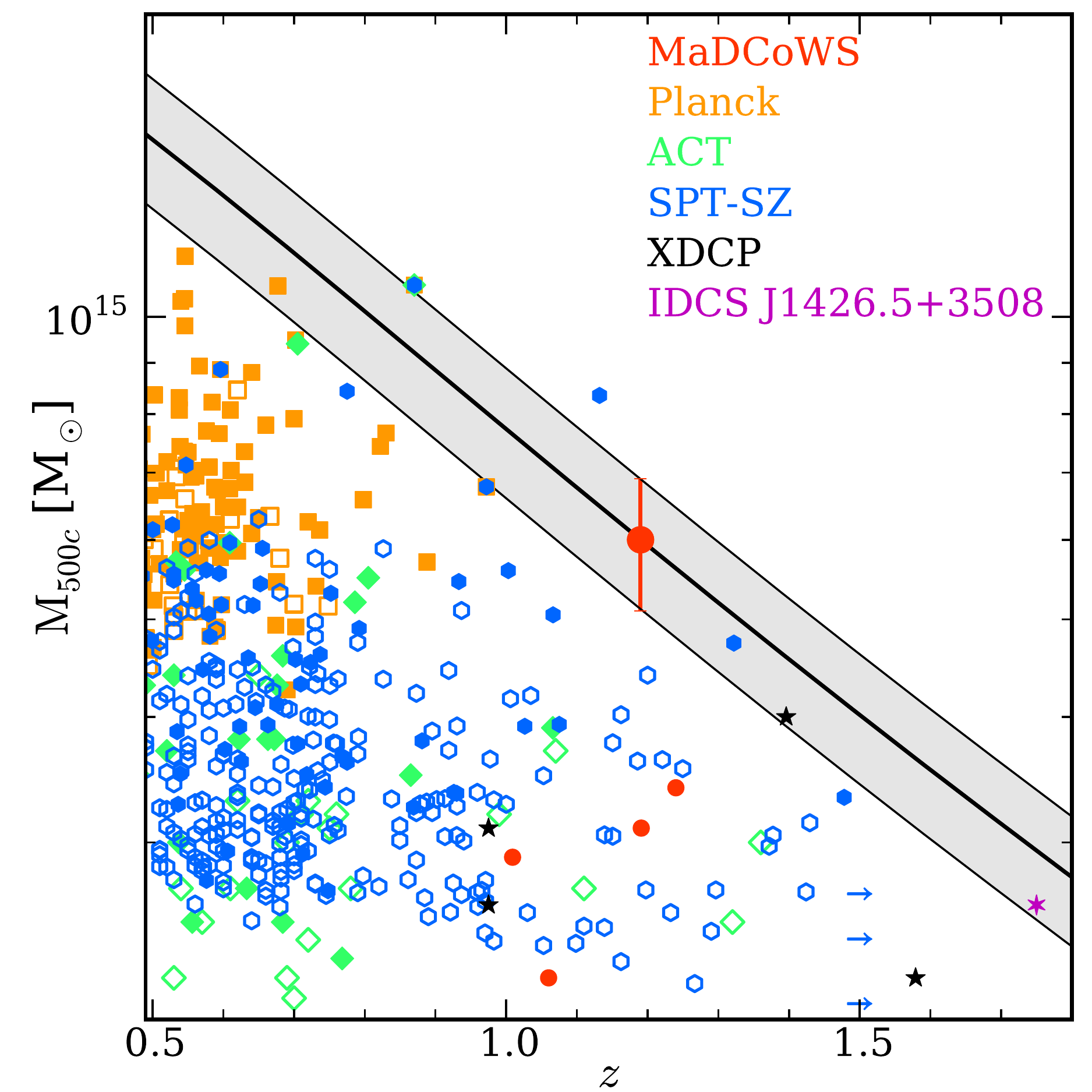}
\caption{Comparison in the mass-redshift plane of \target\ (large red circle with error bars) with other \madcows\ clusters 
\citep[red circles,][]{brodwin2015}, with
  clusters from the \planck\ \citep[orange squares,][]{planck2014xxix}, ACT
  \citep[green diamonds,][]{marriage2011,hasselfield2013}, SPT \citep[blue hexagons,][]{bleem2015}, and
  XDCP \citep[black stars,][]{fassbender2011} surveys, and with IDCS J1426.5+3508 \citep[purple six-pointed star,][]{brodwin2012}. We plot the mass from \planck\ for clusters detected in multiple surveys. 
We use filled symbols for clusters with published spectroscopic redshifts, and open symbols those with photometric redshifts. 
We make the assumption that \planck\ redshifts are spectroscopic in instances where the type of redshift is unclear.
Arrows denote lower limits on SPT photometric redshifts. For XDCP J0044.0-2033 ($z=1.58$) we plot the updated mass from \citet{tozzi2015} that uses the \citet{vikhlinin2009} scaling relation.  
The black line is a curve of
constant comoving number density for a \citet{tinker2008} mass function;
the shaded region 
indicates the corresponding extension of the 1$\sigma$ error bars. }
\label{fig:massz}
\end{figure}

In this paper we have presented confirmation of a massive galaxy
cluster at $z=\targz$. Originally identified by the \madcows\ project,
the cluster 
\target\ has a mass of \mfivec$=(6.0\pm 0.9)\times10^{14}$
\msun\ [\mtwom$=(1.1\pm 0.2)\times10^{15}$ \msun], making it the most
massive confirmed galaxy cluster at $z\ge1.15$ identified by any
technique. Figure \ref{fig:massz} illustrates the position of this
cluster in the mass -- redshift plane compared to a selection of
recent wide-area cluster surveys.  The solid black curve in this
Figure is a curve of constant co-moving number density, highlighting
that there are few clusters over this entire redshift interval as rare
as \target.  The only more massive cluster known at $z>1$ is SPT-CL
J2106-5844 \citep[$z=1.13$, \mtwom$=(1.27\pm0.21)\times 10^{15}
h_{70}^{-1}$ \msun;][]{foley2011}.
We also include in this Figure IDCS J1426.5+3508 ($z=1.75$), as it is the closest progenitor analog to \target\ at $z>1.5$.

The existence of \target\ is not in tension with the \lcdm\
paradigm, but such clusters are expected to be extremely rare. We use
the halo mass function code {\tt hmf} from
\citet{murray2013}\footnote{See also
  \url{https://github.com/steven-murray/hmf}.}  with a
\citet{tinker2008} mass function to calculate the expected number of
such clusters.  For WMAP9 and \planck\ \citep{planck2014xvi}
cosmologies, there are predicted to only be $\sim 3$ or $\sim 7$
clusters this massive over the full sky at $z\ga 1.19$, respectively,
and only $\sim$ 1--2 within our SDSS survey area.  The discovery of
this cluster highlights the potential of wide area cluster surveys like
\madcows\ to identify such extreme systems, which are natural targets
for a range of cosmological and evolutionary investigations. Our
ongoing Cycle 11 \spitzer\ program (PID 11080, PI Gonzalez), which
targets $\sim1750$ additional \madcows\ candidates drawn from the full
extragalactic sky, promises to enable construction of a {\it sample}
of comparably rich galaxy clusters at this epoch.

\acknowledgements 

Support for this research was provided by NASA through \spitzer\ GO
program 90177, ADAP grant NNX12AE15G, and NASA Exoplanet Science
Institute grants 1461527 and 1486927.  The work by SAS at LLNL was
performed under the auspices of the U.~S.~Department of Energy under
Contract No. W-7405-ENG-48.

Support for CARMA construction was derived from the Gordon and Betty
Moore Foundation; the Kenneth T. and Eileen L. Norris Foundation; the
James S. McDonnell Foundation; the Associates of the California
Institute of Technology; the University of Chicago; the states of
California, Illinois, and Maryland; and the National Science
Foundation. CARMA development and operations were supported by 
NSF under a cooperative agreement and by the
CARMA partner universities; the work at Chicago was supported by NSF
grant AST-1140019. Additional support was provided by
PHY-0114422. This publication makes use of data products from the {\it
  Wide-field Infrared Survey Explorer}, 
a joint project of
the University of California, Los Angeles, and the Jet Propulsion
Laboratory/California Institute of Technology, funded by 
NASA. 
This work is based in part on
observations made with the \spitzer\ {\it Space Telescope}, which is
operated by the Jet Propulsion Laboratory, California Institute of
Technology under a contract with NASA. This work is based in part on
data obtained at the W. M. Keck and Gemini Observatories.
The authors wish
to recognize and acknowledge the very significant cultural role and
reverence that the summit of Mauna Kea has always had within the
indigenous Hawaiian community.  We are most fortunate to have the
opportunity to conduct observations from this mountain.

{\it Facility:}\facility{\wise}, \facility{\spitzer~(IRAC)}, \facility{CARMA}, \facility{Keck:I~(MOSFIRE)}, \facility{Keck:II~(DEIMOS)} \facility{Gemini:Gillett~(GMOS)}

\bibliographystyle{apj}
\bibliography{ms}

\end{document}